\documentclass[12pt]{article}
\begin{document}
\date{Today}
\title{{\bf{\Large Path-integral action of a particle in the noncommutative phase-space }}}

\author{
{\bf {\normalsize Sunandan Gangopadhyay}$^{a,b}
$\thanks{sunandan.gangopadhyay@gmail.com, sunandan@associates.iucaa.in}},
{\bf {\normalsize Aslam Halder }$^{c}$\thanks{aslamhalder.phy@gmail.com}}\\[0.2cm]
$^{a}$ {\normalsize Department of Physics, West Bengal State University,
Barasat, Kolkata 700126, India   }\\[0.2cm]
$^{b}${\normalsize Visiting Associate in Inter University Centre for Astronomy $\&$ Astrophysics (IUCAA),}\\
{\normalsize Pune, India}\\[0.3cm]
$^{c}$ {\normalsize Kolorah H.A.W Institution, Kolorah, Howrah-711411, India}\\[0.2cm]
}
\date{}

\maketitle
\noindent
In this paper we construct a path integral formulation of quantum mechanics on noncommutative phase-space. We first map the system to an equivalent system on the noncommutative plane. Then by applying the formalism of representing a quantum system in the space of Hilbert-Schmidt operators acting on noncommutative configuration space, the path integral action of a particle is derived. It is observed that the action has a similar form to that of a particle in a magnetic field in the noncommutative plane. From this action the energy spectrum is obtained for the free particle and the harmonic oscillator potential. We also show that the nonlocal nature (in time) of the action yields a second class constrained system from which the noncommutative Heisenberg algebra can be recovered.

\vskip 1cm

Physics on noncommutative (NC) spacetime has attracted a lot of interest in recent times. This idea, which was introduced originally by Snyder \cite{snyder} to regulate the divergences arising in quantum field theories, came up from the analysis of matrix models in string theory \cite{sw}. Subsequently, lot of activities took place in quantum mechanics on NC space \cite{duval}-\cite{Modak} and quantum field theories on such space \cite{szabo}-\cite{carroll}.

Attention has also been paid recently to the formal and interpretational aspects of NC quantum mechanics \cite{fgs}. This played a key role in developing the path integral formalism of NC quantum mechanics \cite{sgfgs} using coherent states. Despite this development, an open question which remained was the derivation of the action in noncommutative phase-space (NCPS).

In this paper, we obtain the path integral action for a particle
in an arbitrary potential in NCPS. To do so we first map the system to an equivalent system on the NC space. Then by using the formalism in \cite{fgs}, where a quantum system can be represented in the space of Hilbert-Schmidt operators, we obtain the path integral action. Interestingly, the action is nonlocal in time and has a similar form to the action for a particle in an external magnetic field on the NC plane. We then obtain the equation of motion for the particle and compute the ground state energy of the particle for two different cases, namely, the free particle and in the presence of a harmonic
oscillator potential. We further show that the nonlocality in time of the action leads to a second class constrained system which upon quantization yields the NC Heisenberg algebra. 

We start with the following NCPS algebra
\begin{eqnarray}
[\hat{x}_i , \hat{x}_j]=i\theta\epsilon_{ij}~,~[\hat{x}_i , \hat{p}_j]=i\tilde{\hbar}\delta_{ij}~,
~[\hat{p}_i , \hat{p}_j]=i\bar{\theta}\epsilon_{ij}~;(i, j=1, 2) 
\label{nc_alg 1}
\end{eqnarray}
where $\theta$ and $\bar{\theta}$ are the spatial and momentum NC parameters and $\tilde{\hbar}=\hbar(1+\frac{\theta\bar{\theta}}{4\hbar^2})$ is the effective Planck's constant. 
The Hamiltonian  for a particle in the presence of an arbitrary potential $V(\hat{x},\hat{y})$ on NCPS reads
\begin{eqnarray}
\hat{H}=\sum_{i=1}^{2}\frac{\hat{p}_{i}^2}{2m}+V(\hat{x},\hat{y})~.                                                                   
\label{hamil}
\end{eqnarray}
Clearly the formalism developed in \cite{fgs} to study quantum systems on the NC plane cannot be applied at this stage. Hence we map the NC momenta  
 to their equivalent commutative form using a set of transformations known as Bopp-shift defined as \cite{mezin}
\begin{eqnarray}
\label{e6a} 
\hat{p}_{i}  =  p_{i}+\frac{1}{2\hbar}\bar{\theta}_{ij}x_{j}~. 
\end{eqnarray}
\begin{eqnarray}
\label{e6}
\hat{x}_{i}  =   x_{i}-\frac{1}{2\hbar}\theta_{ij}p_{j}~.
\end{eqnarray} 
Note that the spatial coordinates that arise in mapping the NC momenta to their commutative counterpart (using eq.(\ref{e6a})) in eq.(\ref{hamil}) are commutative in nature and therefore we use eq.(\ref{e6}) to map these commutative spatial variables to their equivalent NC form. 
Following this prescription, the Hamiltonian (\ref{hamil}) can be recast in the following form
\begin{eqnarray}
\label{e546a} 
\hat{H}=\frac{1}{2m}\left[a^2(p_{x}^2+p_{y}^2)+\frac{\bar{\theta}^2}{4\hbar^2}(\hat{x}^2+\hat{y}^2)-\frac{a\bar{\theta}}{\hbar}(\hat{x}p_{y}-\hat{y}p_{x})\right]+V(\hat{x},\hat{y})
\end{eqnarray}
where $a=\left(1-\frac{\theta\bar{\theta}}{4\hbar^2}\right)$. The above Hamiltonian involves coordinates and momenta which satisfy the NC Heisenberg algebra
\begin{eqnarray}
[\hat{x}_i , \hat{x}_j]=i\theta\epsilon_{ij}~,~[\hat{x}_i , \hat{p}_j]=i\hbar\delta_{ij}~,
~[\hat{p}_i , \hat{p}_j]=0~;~(i, j=1, 2)~.
\label{nc_alg 1}
\end{eqnarray}
We are now in a position to apply the formalism of NC quantum mechanics \cite{fgs}.

To start with, we give a brief review of this formalism. The first step in this formalism is to consider only the NC algebra between the coordinates $\hat{x}$, $\hat{y}$ which is defined to be the classical configuration space. 
Annihilation and creation operators are then defined by
$\hat b = \frac{1}{\sqrt{2\theta}} (\hat{x}+i\hat{y})$,
$\hat{b}^\dagger =\frac{1}{\sqrt{2\theta}} (\hat{x}-i\hat{y})$
and satisfy the Fock algebra $[\hat{b}, \hat{b}^\dagger ] = 1$. 
The NC configuration space is then 
isomorphic to the boson Fock space
\begin{eqnarray}
\label{e421}
\mathcal{H}_c = \textrm{span}\{ |n\rangle= 
\frac{1}{\sqrt{n!}}(\hat{b}^\dagger)^n |0\rangle\}_{n=0}^{n=\infty}
\end{eqnarray}
where the span is taken over the field of complex numbers.
The Hilbert space
of the NC quantum system is then introduced as 
\begin{eqnarray}
\mathcal{H}_q =\psi(\hat{x},\hat{y}): 
\psi(\hat{x},\hat{y})\in \mathcal{B}
\left(\mathcal{H}_c\right),\;\nonumber\\
{\rm tr_c}(\psi^\dagger(\hat{x},\hat{y})
\psi(\hat{x},\hat{y})) < \infty
\label{e4}
\end{eqnarray}
where ${\rm tr_c}$ denotes the trace over NC 
configuration space and $\mathcal{B}\left(\mathcal{H}_c\right)$ 
the set of bounded operators on $\mathcal{H}_c$. 
This space has a natural inner product and norm 
\begin{eqnarray}
\left(\phi(\hat{x}, \hat{y}), \psi(\hat{x},\hat{y})\right) = 
{\rm tr_c}(\phi(\hat{x}, \hat{x})^\dagger\psi(\hat{x}, \hat{y}))
\label{einner}
\end{eqnarray}
and forms a Hilbert space \cite{hol}. A unitary representation of the NC Heisenberg algebra in terms of operators $\hat{X}$, $\hat{Y}$, $\hat{P}_x$ and $\hat{P}_y$ acting on the states of the quantum Hilbert space 
(\ref{e4}) reads
\begin{eqnarray}
\hat{X}\psi(\hat{x},\hat{y}) &=& \hat{x}\psi(\hat{x},\hat{y})\qquad,\qquad
~\hat{Y}\psi(\hat{x},\hat{y}) = \hat{y}\psi(\hat{x},\hat{y})~,\nonumber\\
\hat{P}_x\psi(\hat{x},\hat{y}) &=& \frac{\hbar}{\theta}[\hat{y},\psi(\hat{x},\hat{y})]\quad,\quad
\hat{P}_y\psi(\hat{x},\hat{y}) = -\frac{\hbar}{\theta}[\hat{x},\psi(\hat{x},\hat{y})]~.
\label{eaction}
\end{eqnarray}
The minimal uncertainty states on NC 
configuration space 
are well known to be the normalized coherent states \cite{klaud}
\begin{eqnarray}
\label{ecs} 
|z\rangle = e^{-z\bar{z}/2}e^{z b^{\dagger}} |0\rangle
\end{eqnarray}
where, $z=\frac{1}{\sqrt{2\theta}}\left(x+iy\right)$ 
is a dimensionless complex number. 
Corresponding to these states we can construct a state 
(operator) in quantum Hilbert space as follows
\begin{eqnarray}
|z, \bar{z} )=\frac{1}{\sqrt{\theta}}|z\rangle\langle z|~.
\label{ecsqh}
\end{eqnarray}
These states have the property
\begin{eqnarray}
\hat{B}|z, \bar{z})=z|z, \bar{z})~;~\hat{B}=\frac{1}{\sqrt{2\theta}}(\hat{X}+i\hat{Y})~.
\label{ep1}
\end{eqnarray}
The `position' representation of a state 
$|\psi)=\psi(\hat{x},\hat{y})$ can be constructed as
\begin{eqnarray}
(z, \bar{z}|\psi)=\frac{1}{\sqrt\theta}tr_{c}
(|z\rangle\langle z| \psi(\hat{x},\hat{y}))\nonumber\\
=\frac{1}{\sqrt\theta}\langle z|\psi(\hat{x},\hat{y})|z\rangle~.
\label{posrep}
\end{eqnarray}
The wave-function of a ``free particle" on the NC plane is given by \cite{sgfgs}
\begin{eqnarray}
(z, \bar{z}|p)=\frac{1}{\sqrt{2\pi\hbar^{2}}}
e^{-\frac{\theta}{4\hbar^{2}}\bar{p}p}
e^{i\sqrt{\frac{\theta}{2\hbar^{2}}}(p\bar{z}+\bar{p}z)}
\label{eg3}
\end{eqnarray}
where $|p)$ are the normalised momentum eigenstates given by
\begin{eqnarray}
|p)&=&\sqrt{\frac{\theta}{2\pi\hbar^{2}}}e^{i\sqrt{\frac{\theta}{2\hbar^2}}
(\bar{p}b+pb^\dagger)}~;~\hat{P}_i |p)=p_i |p)\\
p_x&=&{\mbox Re}\,p~,~p_y={\mbox Im}\,p\nonumber
\label{eg}
\end{eqnarray}
satisfying the completeness relation
\begin{eqnarray}
\int d^{2}p~|p)(p|=1_{Q}~.
\label{eg5}
\end{eqnarray}
It is also a simple matter to prove that the completeness relations for the position eigenstates $|z,\bar{z})$ reads
\begin{eqnarray}
\int \frac{dzd\bar{z}}{\pi}~|z, \bar{z})\star(z, \bar{z}|=1_{Q}
\label{eg6}
\end{eqnarray}
where the star product between two functions 
$f(z, \bar{z})$ and $g(z, \bar{z})$ is defined as
\begin{eqnarray}
f(z, \bar{z})\star g(z, \bar{z})=f(z, \bar{z})
e^{\stackrel{\leftarrow}{\partial_{\bar{z}}}
\stackrel{\rightarrow}{\partial_z}} g(z, \bar{z})~.
\label{eg7}
\end{eqnarray}
One can therefore see that the position representation of the NC
system maps quite naturally to the Voros plane.
With the above formalism in hand, we now proceed to develop the
path integral representation for the propagation kernel
on the two dimensional NC plane. Upto a constant factor this reads
\begin{eqnarray}
(z_f, t_f|z_0, t_0)=\lim_{n\rightarrow\infty}\int
\prod_{j=1}^{n}(dz_{j}d\bar{z}_{j})~(z_f, t_f|z_n, t_n)\star_n
(z_n, t_n|....|z_1, t_1)\star_1(z_1, t_1|z_0, t_0)~.
\label{pint1}
\end{eqnarray}
With the Hamiltonian acting on the quantum Hilbert space
\begin{eqnarray}
\label{e546ab} 
\hat{H}=\frac{1}{2m}\left[a^2(P_{x}^2+P_{y}^2)+\frac{\theta \bar{\theta}^2}{2\hbar^2} B^{\dagger}B+\frac{a\bar{\theta}}{\hbar}\sqrt{\frac{\theta}{2}}\left\{i(B^{\dagger}-B)P_{x}-(B+B^{\dagger})P_{y}\right\}+\frac{\theta\bar{\theta}^2}{4\hbar^2}\right]\nonumber\\
+V(\hat{X},\hat{Y})
\end{eqnarray}
we now compute the propagator over a small segment in the
above path integral (\ref{pint1}). With the help of eq.(s) (\ref{eg5}) and (\ref{eg3}), we have
\begin{eqnarray}
(z_{j+1}, t_{j+1}|z_j, t_j)&=&(z_{j+1}|e^{-\frac{i}{\hbar}\hat{H}\tau}|z_j)\nonumber\\
&=&(z_{j+1}|1-\frac{i}{\hbar}\hat{H}\tau +O(\tau^2)|z_j)\nonumber\\
&=&\int_{-\infty}^{+\infty}d^{2}p_j~e^{-\frac{\theta}{2\hbar^{2}}\bar{p}_j p_{j}}
e^{i\sqrt{\frac{\theta}{2\hbar^{2}}}\left[p_{j}(\bar{z}_{j+1}-\bar{z}_{j})+\bar{p}_{j}(z_{j+1}-z_{j})\right]}\nonumber\\
&&\times e^{-\frac{i}{\hbar}\tau[\frac{a^2\bar{p}_j p_{j}}{2m}+\frac{\theta\bar{\theta}^2}{4m\hbar^2}\bar{z}_{j+1}z_{j}
+\frac{ia\bar{\theta}}{2m\hbar}\sqrt{\frac{\theta}{2}}(p_j \bar{z}_{j+1}-\bar{p}_j z_j)+V(\bar{z}_{j+1},z_{j})]}+O(\tau^2)~.\nonumber\\
\label{pint2}
\end{eqnarray} 
Substituting the above expression in eq.(\ref{pint1}) and computing the star products explicitly, we obtain (apart from a constant factor)
\begin{eqnarray}
(z_f, t_f|z_0, t_0)=&&\lim_{n\rightarrow\infty}\int \prod_{j=1}^{n} (dz_{j}d\bar{z}_{j})
\prod_{j=0}^{n}d^{2}p_{j}\nonumber\\
&&\exp\sum_{j=0}^{n}\left[\frac{i}{\hbar}\sqrt{\frac{\theta}{2}}\left\{p_{j}\left(\bar{z}_{j+1}-\bar{z}_{j}\right)+\bar{p}_{j}\left(z_{j+1}-z_{j}\right)\right\}+\alpha p_{j}\bar{p}_{j}\right.\nonumber\\
&&\left. +\frac{a\tau\bar{\theta}}{2m\hbar^2}\sqrt{\frac{\theta}{2}}\left(\bar{z}_{j+1}p_{j}-z_{j}\bar{p}_{j}\right)-\frac{i\tau\theta\bar{\theta}^2}{4m\hbar^3}\bar{z}_{j+1}z_{j}-\frac{i}{\hbar}\tau V(\bar{z}_{j+1},z_{j})\right]\nonumber\\
&&~~~~~~~~~~~~~~~~\times\exp\left(\frac{\theta}{2\hbar^{2}}\sum_{j=0}^{n-1}p_{j+1}\bar{p}_{j}\right)
\label{pint3}
\end{eqnarray} 
where $\alpha=-\left(\frac{i\tau a^2}{2m\hbar}+\frac{\theta}{2\hbar^{2}}\right)$.
Making the identification $p_{n+1}=p_{0}$, 
a simple inspection shows that the propagation kernel can be recast in the following convenient form
\begin{eqnarray}
(z_f, t_f|z_0, t_0)
=&&\lim_{n\rightarrow\infty}\int \prod_{j=1}^{n} (dz_{j}d\bar{z}_{j})
\prod_{j=0}^{n}d^{2}\tilde{p}_{j}\exp\left(-\vec{\partial}_{z_{f}}\vec{\partial}_{\bar{z}_{0}}\right)\nonumber\\
&&\times\exp\left(\sum_{j=0}^{n}\left[\bar{\tilde{p}}_{j}A\tilde{p}_{j}-BA^{-1}C-\left(\frac{i\tau\theta\bar{\theta}^2}{4m\hbar^3}\right)\bar{z}_{j+1}z_{j}-\frac{i}{\hbar}\tau V(\bar{z}_{j+1},z_{j})\right]\right)\nonumber\\
\label{matrix}
\end{eqnarray}
where
\begin{eqnarray}
\label{Amatrix}
A&=&\frac{\tau}{\hbar}\left(-\frac{ia^2}{2m}+\frac{\theta}{2\hbar}\partial_{\tau}\right)\\
B&=&\sqrt{\frac{\theta}{2\hbar^2}}\left(i+\frac{\tau a\bar{\theta}}{2m\hbar}\right)\bar{z}_{j+1}-i\sqrt{\frac{\theta}{2\hbar^2}}\bar{z}_{j}\\
C&=&i\sqrt{\frac{\theta}{2\hbar^2}}z_{j+1}-\sqrt{\frac{\theta}{2\hbar^2}}\left(i+\frac{\tau a\bar{\theta}}{2m\hbar}\right)z_{j}.
\end{eqnarray}
One can carry out the momentum integral easily to obtain 
\begin{eqnarray}
(z_f, t_f|z_0, t_0)=&&\lim_{n\rightarrow\infty}N\int\prod_{j=1}^{n}(dz_{j}d\bar{z}_{j})\exp\left(-\vec{\partial}_{z_{f}}\vec{\partial}_{\bar{z}_{0}}\right)\nonumber\\
&&\times\exp\left[\left(\frac{im\theta}{\tau\hbar a^2}\right)
\left\{\tau \dot{\bar{z}_{j}}-\frac{i\tau a\bar{\theta}}{2m\hbar}\bar{z}_{j}\right\}\right.\nonumber\\
&&\left.\times\left(1+\frac{im\theta}{\hbar a^2}\partial_{\tau}\right)^{-1}\left\{\tau \bar{z}_{j}+\frac{i\tau a\bar{\theta}}{2m\hbar}z_{j}\right\}\right]\nonumber\\
&&\times\exp\left[-\left(\frac{i\tau\theta\bar{\theta}}{4m\hbar^3}\right)\bar{z}_{j}z_{j}-\frac{i}{\hbar}\tau V(\bar{z}_{j}z_{j})+O(\tau^{2})\right].
\label{pintegral1}
\end{eqnarray} 
In the above step we have used the fact that $z_{j}=z(j\tau)$
and $z_{j+1}-z_{j}=\tau\dot{z}(j\tau)+O(\tau^{2})$. Taking the 
limit $\tau\rightarrow 0$, we finally
arrive at the path integral representation of the propagator
\begin{eqnarray}
(z_f, t_f|z_0, t_0)&=&N\exp\left(-\vec{\partial}_{z_{f}}\vec{\partial}_{\bar{z}_{0}}\right)\int_{z(t_0)=z_0}^{z(t_f)=z_f }\mathcal{D}z\mathcal{D}\bar{z}
\exp({\frac{i}{\hbar}S})
\label{pintegral3}
\end{eqnarray} 
where $S$ is the action given by
\begin{eqnarray}
S=\int_{t_{0}}^{t_{f}}dt~\theta \left[\frac{m}{a^2}\left\{\dot{\bar{z}}(t)-\frac{ia\bar{\theta}}{2m\hbar}\bar{z}(t)\right\}\left(1+\frac{im\theta}{\hbar a^2}
\partial_{t}\right)^{-1}
\left\{\dot{z}(t)+\frac{ia\bar{\theta}}{2m\hbar}z(t)\right\}\right.\nonumber\\
\left.-\frac{\bar{\theta}^2}{4m\hbar^2}\bar{z}(t)z(t)-V(\bar{z}(t),z(t))\right].\nonumber\\
\label{ACTION}
\end{eqnarray}
The above action has a form similar to that of a particle in a magnetic field in the NC plane \cite{SGFS}, \cite{SGFGS2}. 
We now compute the ground state energy for the particle without any interaction and in the presence of a harmonic oscillator
potential $V=\frac{1}{2}m\omega^2(\hat{X}^{2}+\hat{Y}^{2})$ from the above path integral representation of the transition amplitude. For the free particle, the above action reduces to the following form 
\begin{eqnarray}
S=\int_{t_{0}}^{t_{f}}dt~\theta \left[\frac{m}{a^2}\left\{\dot{\bar{z}}(t)-\frac{ia\bar{\theta}}{2m\hbar}\bar{z}(t)\right\}\left(1+\frac{im\theta}{\hbar a^2}
\partial_{t}\right)^{-1}
\left\{\dot{z}(t)+\frac{ia\bar{\theta}}{2m\hbar}z(t)\right\}
-\frac{\bar{\theta}^2}{4m\hbar^2}\bar{z}(t)z(t)\right].
\label{action_ncqm free}
\end{eqnarray}
The equation of motion following from the above action reads
\begin{eqnarray}
\ddot{z}(t)+\frac{ia\bar{\theta}}{m\hbar}\left(1+\frac{a\theta\bar{\theta}}{4\hbar^2}\right)\dot{z}(t)=0~.
\label{free particle}
\end{eqnarray}
Making an ansatz $z(t)\sim e^{i\gamma t}$
leads to the following ground state energy eigenvalues for the particle  
\begin{eqnarray}
\gamma=-\frac{a\bar{\theta}}{m\hbar}\left(1+\frac{a\theta\bar{\theta}}{4\hbar^2}\right), 0~.
\label{enereigval}
\end{eqnarray} 
This ground state energy, computed from the path integral formalism, matches with those obtained by the canonical approach \cite{SGFS}.

In the presence of a harmonic oscillator potential, the action takes the form
\begin{eqnarray}
S=\int_{t_{0}}^{t_{f}}dt~\theta \left[\frac{m}{a^2}\left\{\dot{\bar{z}}(t)-\frac{ia\bar{\theta}}{2m\hbar}\bar{z}(t)\right\}\left(1+\frac{im\theta}{\hbar a^2}
\partial_{t}\right)^{-1}
\left\{\dot{z}(t)+\frac{ia\bar{\theta}}{2m\hbar}z(t)\right\}\right.\nonumber\\
\left.-\left(m\omega^2+\frac{\bar{\theta}^2}{4m\hbar^2}\right)\bar{z}(t)z(t)\right]~.
\label{action_ncqm ho}
\end{eqnarray} 
The equation of motion following from the above action reads
\begin{eqnarray}
\ddot{z}(t)+i\left\{\frac{a\bar{\theta}}{m\hbar}\left(1+\frac{a\theta\bar{\theta}}{4\hbar^2}\right)+\frac{m\omega^2 a^2\theta}{\hbar}\right\}\dot{z}(t)+\omega^2 z(t)=0~.
\label{harsol}
\end{eqnarray} 
Making a similar ansatz as in the free particle case
leads to the following ground state energy eigenvalues for the particle  
\begin{eqnarray}
\gamma=\frac{1}{2}\left[-\left\{\frac{m\omega^{2}a^2\theta}{\hbar}+\frac{a\bar{\theta}}{m\hbar}\left(1+\frac{a\theta\bar{\theta}}{4\hbar^2}\right)\right\}
\pm\sqrt{\left(\frac{m\omega^{2}a^2\theta}{\hbar}+\frac{a\bar{\theta}}{m\hbar}\left(1+\frac{a\theta\bar{\theta}}{4\hbar^2}\right)\right)^2+4\omega^{2}}\right]~.\nonumber\\
\label{energyeig}
\end{eqnarray} 
In the $\bar{\theta}\rightarrow0$ limit, the above expression yields the two frequencies for a particle in a harmonic oscillator potential on the NC plane \cite{sgfgs}.
 
As a consistency check, we quantize this theory to see if
we recover the NC Heisenberg algebra. To do
this we first introduce an auxiliary complex field $\phi$ and
write the transition amplitude as
\begin{eqnarray}
(z_f, t_f|z_0, t_0)&=&N\exp\left(-\vec{\partial}_{z_{f}}\vec{\partial}_{\bar{z}_{0}}\right)\int\mathcal{D}\phi\mathcal{D}\bar{\phi}\int_{z(t_0)=z_0}^{z(t_f)=z_f }\mathcal{D}z\mathcal{D}\bar{z}
\exp({\frac{i}{\hbar}S})
\label{pintegral4}
\end{eqnarray}
where the action $S$ is given by 
\begin{eqnarray}
S=\int_{t_{0}}^{t_{f}}dt \left[-\bar{\phi}\left(\frac{a^2}{m\theta}+\frac{i}{\hbar}\partial_{t}\right)\phi+\phi\left\{\dot{\bar{z}}(t)-\frac{ia\bar{\theta}}{2m\hbar}\bar{z}(t)\right\}+\bar{\phi}\left\{\dot{z}(t)+\frac{ia\bar{\theta}}{2m\hbar}z(t)\right\}\right.\nonumber\\
\left.-\theta\left(m\omega^2+\frac{\bar{\theta}^2}{4m\hbar^2}\right)\bar{z}(t)z(t)\right]~.
\label{action_ncqm 1}
\end{eqnarray} 
It is evident from the form of the action that this is a constrained
system with the following second class constraints
\begin{eqnarray}
&\chi_{1}=p_{\phi_{1}}+\frac{1}{\hbar}\phi_{2}\approx0~, \quad\chi_{2}=p_{\phi_{2}}-\frac{1}{\hbar}\phi_{1}\approx0~,\nonumber\\
&\chi_{3}=p_{x}-\sqrt{\frac{2}{\theta}}\phi_{1}\approx0~,\quad\chi_{4}=p_{y}+\sqrt{\frac{2}{\theta}}\phi_{2}\approx0~.
\label{constr}
\end{eqnarray} 
Introducing the Dirac bracket and replacing $\{. ,.\}_{DB}\rightarrow\frac{1}{i\hbar}[. ,.]$ yield the following
NC algebra
\begin{eqnarray}
[\hat{x}_i , \hat{x}_j]=\theta\epsilon_{ij}~,~[\hat{x}_i , \hat{p}_j]=i\hbar\delta_{ij}~,
~[\hat{p}_i , \hat{p}_j]=0~;~(i, j=1, 2).
\label{nc_alg}
\end{eqnarray} 
In summary, we have developed the path integral
representation of the propagation kernel for a particle in the presence of an arbitrary potential moving in the the noncommutative phase-space using the formalism representing quantum system in the space of Hilbert-Schmidt operators acting in the noncommutative configuration space. From this path integral formulation, we have obtained the action
for the particle and has a structure similar to that of a particle in a magnetic field in the noncommutative plane. This is one of the main results in our paper. The equation of motion of the particle is obtained from this action and is used
to compute the ground state energy of the particle in two different cases, namely, that of a free particle and a harmonic oscillator. The results, obtained here, are in conformity with the results obtained by other methods \cite{SGFS}, \cite{ah}. We also show that the nonlocality in time of the action leads to a second class constrained system from which the noncommutative Heisenberg algebra can be recovered upon quantizing the theory.

\section*{Acknowledgements}
S.G. acknowledges the support by DST SERB under Start Up Research Grant (Young Scientist), File No.YSS/2014/000180.

\end{document}